# Rapid 3D Multiparametric Mapping of Brain Metastases with Deep Learning-Based Phase-Sensitive MR Fingerprinting

Victoria Y. Yu, Kathryn R. Tringale, Ricardo Otazo, and Ouri Cohen

***Abstract*—In MR fingerprinting (MRF) reconstruction, measured data is pattern-matched to simulated signals to extract quantitative tissue parameters. A critical drawback to this approach is the exponentially increasing compute time for mapping of multiple parameters. Previously, a deep learning (DL) reconstruction method called DRONE was shown to overcome this constraint by mapping the magnitude time-series signal to the underlying tissue parameters. However, relaxometry from magnitude images is susceptible to errors arising from ambiguities in the zero crossing of the signal or the non-zero noise mean. The aim of this study is to develop rapid acquisition and quantification methods to enable accurate multiparametric tissue mapping from complex data. An optimized EPI based MRF sequence is developed along with a novel phase-sensitive DL quantification allowing the use of real-valued neural networks to reconstruct complex measured data and providing an additional quantitative map of the phase. Phantom experiments demonstrate the accuracy of the proposed approach. A comparison to previous DRONE methods in a healthy subject shows improved fidelity to known T1 and T2 values for the phase-sensitive approach. By processing the estimated phase map with conventional quantitative susceptibility mapping algorithms, we demonstrate the feasibility of simultaneous quantification of proton density, T1, T2, transmitter B1+ field and the quantitative susceptibility maps. In vivo experiments in a healthy volunteer and a subject with metastatic brain cancer are used to illustrate potential applications of this technology for treatment response assessment and tumor characterization.***

## I. INTRODUCTION

MAGNETIC resonance imaging (MRI) is widely used in cancer imaging because it provides multiple image contrast and superior delineation of soft tissue lesions compared to computed tomography (CT)[1], [2] . The versatility of MRI is a result of the dependence of the MR signal on various tissue and instrumental parameters. In a conventional MRI, the effects of these parameters are not generally considered and clinical interpretation is based on subjective assessment of signal intensity changes in the images. While techniques for mapping tissue parameters have long existed [3], [4], the required scan time was historically prohibitive for routine clinical use.

MR fingerprinting (MRF) is a recent technique [5] that enables rapid quantitative mapping of multiple tissue parameters simultaneously. The tissue maps are obtained by dynamically varying the acquisition parameters to induce a differential signal evolution for different tissue types. The MRF framework is flexible and allows multiple pulse sequences and k-space sampling schemes to be used [6], [7]. MRF was originally proposed using heavily undersampled time frames, which requires many time frames for robust parametric mapping. The ability of echo planar imaging (EPI) to traverse k-space in a single shot offers a powerful alternative that is less susceptible to undersampling artifacts and thus requires fewer time frames [8]–[11].

To generate tissue parameter maps, the measured signal is pattern-matched to a pre-computed dictionary of simulated signal magnetizations. Because the dictionary grows exponentially with the number of parameters, dictionary matching is constrained to a small number of simultaneously estimated parameters, severely limiting the potential of this approach. Previously, a deep learning method named Deep RecOnstruction Network (DRONE) [12] was introduced to overcome this constraint by using a neural network to map the magnitude signal to the underlying tissue parameters. The DRONE network was trained on a simulated sparse dataset, bypassing the limits imposed by the exponentially growing dictionary. Once trained, the DRONE reconstruction was nearly 5000-fold faster and provided continuous-valued tissue parameter maps.

In its original implementation [12], DRONE used a real-valued network and magnitude images to quantify T1 and T2 alone. A drawback of this approach is that relaxometry from magnitude images is susceptible to errors due to, for example, ambiguities in the zero-crossing of the signal or the non-zero noise mean [13]. A simple splitting of the image into real and imaginary components blurs the relationship between the components and can also lead to erroneous results. Some have proposed using complex-valued networks and activation

This research was funded in part by NIH/NCI Cancer Center Support Grant P30-CA008748 and 1R37CA262662-01A1.
Victoria Y. Yu is with Memorial Sloan Kettering Cancer Center, New York, NY 10025 USA (e-mail: YuV@mskcc.org).
Kathryn R. Tringale is with Memorial Sloan Kettering Cancer Center, New York, NY 10025 USA (e-mail: tringalk@mskcc.org).
Ricardo Otazo is with Memorial Sloan Kettering Cancer Center, New York, NY 10025 USA (e-mail: otazotoj@mskcc.org).
Ouri Cohen (corresponding author) is with Memorial Sloan Kettering Cancer Center, New York, NY 10025 USA (e-mail: coheno1@mskcc.org).

functions [14] at the cost of increasing complexity in network training. Additionally, complex operations and optimization algorithms are poorly supported by deep learning software frameworks (e.g. PyTorch) thereby requiring the use of custom, application-specific code. Of note, the phase of the MR signal contains useful information and techniques such as quantitative susceptibility mapping (QSM) [15] exploit this to obtain valuable diagnostic data.

This work proposes to combine an optimized MRF-EPI pulse sequence with a simple modification of DRONE to enable reconstruction of complex data using real-valued networks while simultaneously obtaining additional, useful phase maps. By processing the phase maps with a QSM pipeline [16], this also demonstrates the feasibility of obtaining susceptibility data from the same acquisition. Inspired by other phase-sensitive reconstruction methods [17]–[19], the quantification method is called 'Phase-Sensitive DRONE' (PS-DRONE). We characterize the accuracy of the tissue maps reconstructed with our method in the official ISMRM NIST phantom [20] and demonstrate the clinical utility in a healthy human subject and a patient with brain metastases.

The rest of this paper is structured as follows. Section II presents the theoretical background and formulation of the proposed approach. Section III describes the acquisition and reconstruction methods. Section IV presents simulation and experimental results followed by discussion (section V) and conclusion (section VI).

## II. THEORY

MRI spin dynamics are governed by the Bloch equations [21]. Extended Phase Graphs (EPG) [22] offer a simple yet powerful framework for solving the Bloch equations and modeling MRI pulse sequences. We briefly review the EPG formalism and introduce the proposed modification to enable phase estimation. The subsequent EPG description largely follows the formalism introduced in Weigel et al. [22]. In the following, bold letters denote vectors or matrices, * is the complex conjugate and $\mathbf{X}^T$ is the transpose of $\mathbf{X}$.

Let $\mathbf{M}(\mathbf{r})$ be the magnetization of a single isochromat with magnitude $M = |\mathbf{M}|$ and components ($M_x$, $M_y$, $M_z$). The equations of motion under the effect of a gradient $\mathbf{G}$ are

$$M_x(\mathbf{r}) = M \cos(\mathbf{kr}) \quad (1)$$
$$M_y(\mathbf{r}) = M \sin(\mathbf{kr}) \quad (2)$$

where $\mathbf{k}$ is the angular wave vector is given by

$$\mathbf{k}(t) = \gamma \int_0^t \mathbf{G}(t')dt' \quad (3)$$

The transverse magnetization can be represented more compactly with a complex representation ($i = \sqrt{-1}$)

$$M_+(\mathbf{r}) = M_x(\mathbf{r}) + iM_y(\mathbf{r}) = Me^{i\mathbf{kr}} \quad (1)$$
$$M_-(\mathbf{r}) = M_x(\mathbf{r}) - iM_y(\mathbf{r}) = Me^{-i\mathbf{kr}} \quad (2)$$

The net magnetization is the sum (integral, in the continuous limit) of all isochromats along the gradient over a volume V. A macroscopic sample will have isochromats with different spatial frequencies which naturally lends itself to a Fourier decomposition:

$$\tilde{F}_+(\mathbf{k}) = \int_V M_+(\mathbf{r})e^{-i\mathbf{kr}}d^3\mathbf{r} \quad (3)$$

$$\tilde{F}_-(\mathbf{k}) = \int_V M_-(\mathbf{r})e^{-i\mathbf{kr}}d^3\mathbf{r} \quad (4)$$

$$\tilde{Z}(\mathbf{k}) = \int_V M_z(\mathbf{r})e^{-i\mathbf{kr}}d^3\mathbf{r} \quad (5)$$

Equations (3)-(5) are called the "configuration states" and offer a compact way of representing the state of the net magnetization. In a typical MRI experiment, the magnetization is subject to RF excitation pulses, spin relaxation and gradient dephasing. Given an excitation flip angle α and RF phase β, we define the operator $\mathbf{T}$:

$$\mathbf{T}(\alpha, \beta) = \begin{bmatrix} \cos^2\frac{\alpha}{2} & e^{i2\beta}\sin^2\frac{\alpha}{2} & -ie^{i\beta}\sin\alpha \\ e^{-i2\beta}\sin^2\frac{\alpha}{2} & \cos^2\frac{\alpha}{2} & ie^{-i\beta}\sin\alpha \\ -\frac{i}{2}e^{-i\beta}\sin\alpha & \frac{i}{2}e^{i\beta}\sin\alpha & \cos\alpha \end{bmatrix} \quad (6)$$

Given relaxation parameters T1 and T2 and an equilibrium magnetization $M_0$, we define the operators $\mathbf{E}$, $\mathbf{M}$:

$$\mathbf{E}(t; T1, T2) = \begin{bmatrix} e^{-t/T2} & 0 & 0 \\ 0 & e^{-t/T2} & 0 \\ 0 & 0 & e^{-t/T1} \end{bmatrix} \quad (7)$$

$$\mathbf{M}(t; M_0, T1) = \begin{bmatrix} 0 \\ 0 \\ M_0(1 - e^{-t/T1}) \end{bmatrix} \quad (8)$$

Noting that $(\tilde{F}_+)^* = \tilde{F}_-$ and switching to a discrete representation $\mathbf{F} = [\tilde{F}_k, \tilde{F}_{-k}^*, \tilde{Z}_k]^T$, the effect of RF pulses on the configuration states can be modeled by application of the $\mathbf{T}$ operator: $\mathbf{F}(t^+) = \mathbf{T}(\alpha, \beta)\mathbf{F}(t^-)$ while the effect of spin relaxation can be modeled by the $\mathbf{E}$ and $\mathbf{M}$ operators: $\mathbf{F}(t^+) = \mathbf{E}(\tau)\mathbf{F}(t^-)$ for k≠0 and $\mathbf{F}(t^+) = \mathbf{E}(\tau)\mathbf{F}(t^-) + \mathbf{M}(\tau)$ for k=0. Finally, gradient dephasing is modeled through a shift operator $\mathbf{S}(\Delta k)$ where $\Delta k = \gamma \int_{t'=0}^{t} \mathbf{G}(t')dt'$:

$$\tilde{F}_k \rightarrow \tilde{F}_{k+\Delta k} \quad (9)$$

$$\tilde{Z}_k \rightarrow Z_k \quad (10)$$

The state $\tilde{F}_0$ represents coherent transverse magnetization giving rise to the MR signal. The population of this state determines the magnitude and phase of the resulting signal.

In MRF, flip angles and repetition times (TR) are both varied in each time frame to induce dynamic changes in the magnetization. Given a set of N flip angles $\theta_1 \ldots \theta_N$, N repetition times $TR_1 \ldots TR_N$ and N echo times $TE_1 \ldots TE_N$, and assuming an initial inversion pulse, the resulting state can be calculated by successive applications of the operators defined above:

$$F(t) = S(\Delta k)E(TR_N)T(\theta_n, 90°) \dots$$
$$S(\Delta k)E(TR_1)T(\theta_1, 90°)$$
$$T(180°, 0°)F(t = 0) \quad (11)$$

The MRF signal $S_{MRF}$ is the vector of $\tilde{F}_0$ state values calculated at the appropriate TE for each repetition

$$S_{MRF} = [\tilde{F}_0(TE_1), \dots, \tilde{F}_0(TE_N)] \quad (12)$$

In MRF dictionary matching, the simulated signal defined by Eq. (12) is compared to the measured signal with the best match providing the desired underlying tissue parameters.

Note that in the sequence defined by Eq. (11) the magnetization is initially inverted by a 180° RF pulse and the RF phase of subsequent excitations is 90° out of phase. This constricts the calculated signal to be purely real (zero phase). In an actual scan, magnetic field inhomogeneities arising from, for instance, the main magnetic field or tissue susceptibility sources cause a further, unknown, dephasing that is not accounted for in this model leading to a mismatch between the measured and simulated signals. To account for the unknown phase, we propose a simple modification to the MRF signal:

$$\hat{S}_{MRF} = [\tilde{F}_0(TE_1)e^{i\Phi}, \dots, \tilde{F}_0(TE_N)e^{i\Phi}] \quad (13)$$

Recovery of the unknown phase term $\Phi$ is accomplished by joint estimation of the phase with the other tissue parameters which yields a zero-phase signal. It should be noted that this model assumes a constant phase accrual at each TE. For static susceptibility sources and a constant TE, this assumption is justified because the magnetization is fully dephased in each time point by the gradient dephasing operator $S$ so there are no contributions to the phase from previous time points.

## III. METHODS

### A. Pulse sequence

The proposed MRF-EPI pulse sequence is shown in Fig. 1. The sequence consists of an initial adiabatic inversion pulse followed by a series of fat saturation and excitation pulses (Fig.1a). The fat saturation pulses are chemical shift selective and focused on the lipid protons so do not affect the water signal and are hence not considered in the modeling. The flip angle and TR of the n-th RF excitation pulse are set according to the schedule shown in Fig. 1(b) and (c). The schedule was optimized to maximize discrimination between different tissue types as previously described [23]. In brief, a dictionary $\mathbf{D}$ of signals arising from different tissue parameter values was defined and the matrix $\mathbf{D}^T\mathbf{D}$ calculated. The diagonal entries in the matrix define the correlation between each entry and itself whereas off-diagonal elements define the correlation of a given signal and all other entries in the dictionary. The goal of the optimization was to find the set of flip angles and TRs that minimize the off-diagonal entries of the matrix $\mathbf{D}^T\mathbf{D}$. Following a fixed echo delay, the signal was acquired using an EPI readout.

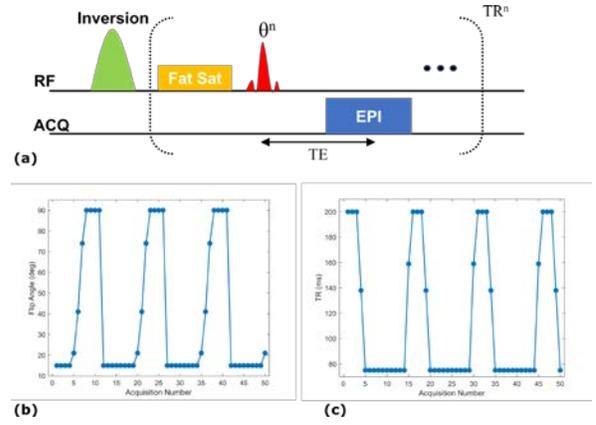

Fig.1. The MRF-EPI pulse sequence and acquisition schedule. (a) Pulse sequence diagram. (b) Optimized schedule of RF excitation pulse flip angles. (c) Optimized schedule of TRs. The 50-point schedule was optimized to maximize the discrimination between different tissue types.

### B. PS-DRONE Quantification

The proposed PS-DRONE method was used to quantify the tissue parameters from the signal acquired with the MRF-EPI pulse sequence. The outline of PS-DRONE is shown in Fig 2. As described in section II, the central idea is that the voxel-wise phase variations in the signal can be jointly estimated with other tissue parameters. Like its predecessor [12], PS-DRONE uses a training dataset of signal magnetizations generated by simulating an MRF acquisition for different tissue parameter values. The tissue parameters include the T1 and T2 relaxation, the transmitter B1-field inhomogeneity as well as the phase term $\Phi$. In the training stage (Fig. 2), the signals calculated with the EPG framework for solving the Bloch equations are multiplied by the complex exponential exp(i$\Phi$) (Eqn. 13) to capture the phase variations in the signal. The complex signal is split into real and imaginary components and the components are concatenated to form the input to the network. In the inference stage, the real and imaginary components of the complex measured signal are separated in the same way and concatenated to then be evaluated by the trained network. Because only real-valued quantities are used in training and inference, standard deep learning frameworks and optimization algorithms can be used. The network is trained to estimate the inverse of the signal phase, i.e. the phase that yields a real-valued signal.

### C. Network Definition and Training

A 4-layer neural network was defined in PyTorch [24]. The network structure included an input, output and two 300-nodes hidden layers. The input layer contained 2N (N=50) nodes, corresponding to the real and imaginary components of the N-points signal. The output layer contained 4 nodes corresponding to the T1, T2, B1 and $\Phi$ tissue maps. The B1 was calculated as a scaling factor of the excitation flip angle. The proton density (PD) of each voxel was calculated as a scaling factor from the reconstructed data using the expression [5]

$$PD = \frac{\mathbf{m}^T\mathbf{d}}{\|\mathbf{d}\|^2} \quad (14)$$

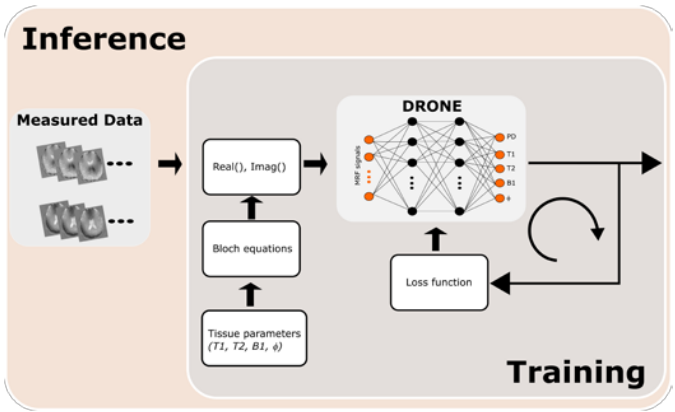

Fig.2. Outline of the PS-DRONE approach. In the training stage, tissue parameters are regularly sampled from the defined ranges and signal magnetization curves are generated using the Bloch equations. The complex signals are split into real and imaginary components and used as input to the DRONE network. Once trained, the network is used to output the tissue parameters from the complex measured data.

Here **m** is the measured signal and **d** is the signal generated by using the tissue parameter values found by the network. The *tanh* activation function was used for the hidden layers with a sigmoid function for the output layer. The network was trained with 80% of a dataset containing 400,000 entries sampled from the tissue parameter ranges using latin hypercube sampling [9] with the remaining 20% used for validation. The following tissue parameter ranges were used: T1 = 1 to 4000 ms, T2 = 1 to 3000 ms, B1 = 0 to 1.5, $\Phi = \pi$ to $3\pi$ rads. A positive range of $\Phi$ values was chosen to avoid truncation of negative values by the sigmoid activation function in the output layer. Because of the periodicity of the complex exponential, any $2\pi$ interval is adequate. The network was trained for 2000 epochs using the ADAM [10] optimizer with an L1 loss with an adaptive learning rate of 0.0001 on an NVIDIA RTX 2080 TI GPU with 11GB of memory. Zero mean Gaussian noise with 1% standard deviation was added to the training data to promote robust learning.

### D. Simulations

The effect of the phase $\Phi$ on the measured signal was simulated for tissues representative of white matter (WM) and gray matter (GM). The WM T1/T2 values were set to 800ms/60ms and the GM values T1/T2 set to 1300ms/80ms [25]. A MRF-EPI acquisition was simulated using the acquisition schedule shown in Fig. 1. Simulations were conducted with the phase set to $\Phi=\pi$ and $\Phi=3\pi/2$ and the resulting real and imaginary signals were compared to the magnitude signal.

### E. MRI

Experiments were conducted on a Signa Premier 3T scanner (GE Healthcare, Waukesha, WI) with a 48-channel head receiver coil. Images were acquired with the MRF-EPI pulse sequence. The acquisition parameters were as follows: partial Fourier factor of ~6/8, parallel imaging acceleration factor R=3, echo time=24 ms, acquisition matrix of 224 × 224 interpolated to 256 × 256, in-plane resolution of 1.1 mm × 1.1 mm with a slice thickness of 5 mm. The MRF acquisition schedule consist of 50 time points with variable flip angle and TR, as shown in Figure 1. The scan time was 5.8 seconds per slice for a total scan time of approximately 3 minutes for whole-brain coverage.

### F. Phantom Testing

The multicompartment ISMRM NIST phantom [20] was used to assess the quantitative accuracy of the T1 and T2 maps quantified with the proposed method. The phantom was scanned with the optimized MRF-EPI sequence and the T1 and T2 maps quantified with PS-DRONE. Regions-of-interest (ROIs) were drawn around each compartment and the mean and standard deviation (SD) calculated and compared to the reference values. The reference values were calculated using gold-standard inversion-recovery and spin-echo sequences [20]. The correlation and the root-mean-square-error (RMSE) between the estimated and reference values was calculated for each parameter.

### G. In Vivo Experiments

#### 1) Quantitative Susceptibility Mapping

The estimated phase maps $\Phi$ obtained were further processed to demonstrate the feasibility of extracting quantitative susceptibility information. A customized QSM pipeline was used (Fig. 3) suitable for the MRF-derived data obtained with PS-DRONE. First, the wrapped, estimated phase maps were unwrapped using the Python Scikit Image library function phase_unwrap() [26]. Because the network estimates the inverse of the measured phase, the polarity of the unwrapped phase was inverted prior to further processing. Unwrapping errors along phase transition boundaries were removed by application of a 3 × 3 median filter. A brain mask was defined by thresholding the PD image followed by image filling and labeling in Matlab (The Mathworks, Natick, MA, USA). Next, the background field was removed using the Laplacian Boundary Value (LBV) method [27] and the susceptibility maps ($\chi$) were calculated using the Morphology Enabled Dipole Inversion (MEDI) toolbox [16], [28]–[30]. The MEDI regularization parameter was set to $\lambda=1000$ and the spherical mean value operator set to 5.

#### 2) Healthy Subject

A healthy, 30-year-old female volunteer was recruited and gave informed consent in accordance with the institutional IRB protocol. The subject was scanned with the MRF-EPI pulse sequence and the data reconstructed with PS-DRONE as described above. For comparison, the data was also reconstructed using standard, magnitude-based, DRONE as well as with PS-DRONE but with the range of phases $\Phi$ set to 0. ROIs were drawn in the WM and GM and the mean and SD of the T1 and T2 values in each region and for each method calculated and compared to values obtained with other methods [31]. The processing time required to generate the tissue maps from the 22-slice volume with the trained neural network was approximately 1 second.

#### 3) Patients with Brain Metastases

A 40-year-old subject with metastatic melanoma was recruited for this study and gave informed consent. The subject was scanned, and the tissue maps reconstructed with PS-DRONE as above. This patient was previously treated with stereotactic radiosurgery, so some tumors had been irradiated whereas others were untreated. Each lesion was segmented into necrotic, edema and tumor regions by a trained radiation oncologist. A healthy contra-lateral region was also demarcated for comparison. The distribution of the T1, T2 and χ values in each tumor region was calculated in comparison to those of the healthy values. The statistical significance of the differences in tissue parameter values for each ROI was calculated using a multi-comparison ANOVA test with Tukey HSD [32] with a significance level set at $p$=0.05.

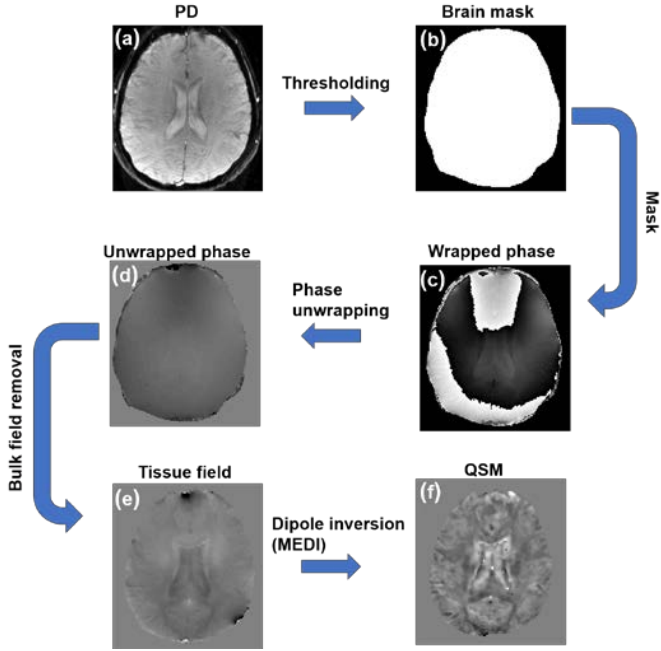

Fig. 3. The pipeline used to derive the QSM maps from the PS-DRONE data. (a) The PD image was thresholded to extract the brain mask (b). The wrapped phase (c) was masked and unwrapped (d) and the bulk field removed using the LBV method to obtain the tissue field (e). Finally, the dipole inversion was applied to the tissue field to yield the susceptibility map shown in (f).

## IV. RESULTS

### A. Simulations

Differences in the WM and GM signals for two different phase values are shown on a two-sided plot in Fig. 4. Differences in the phase led to measurable changes in the real and imaginary components (Figs. 4a-b, 4d-e) supporting the feasibility of inferring the phase from the measured signal. In contrast, the phase information was lost in the magnitude signal (Figs. 4c, 4f).

### B. Phantom Experiments

The mean T1 and T2 values in the phantom reconstructed with PS-DRONE are shown in Fig. 5. Both T1 and T2 were strongly correlated ($R^2$=0.99) with the reference values with a RMSE of 43 ms for T1 (Fig. 5a) and 40 ms for T2 (Fig. 5b). The equation of the best-fit line to the data showed little deviation from linearity for both T1 (2%) and T2 (7%) and minimal bias (T1/T2=3.3ms/11ms).

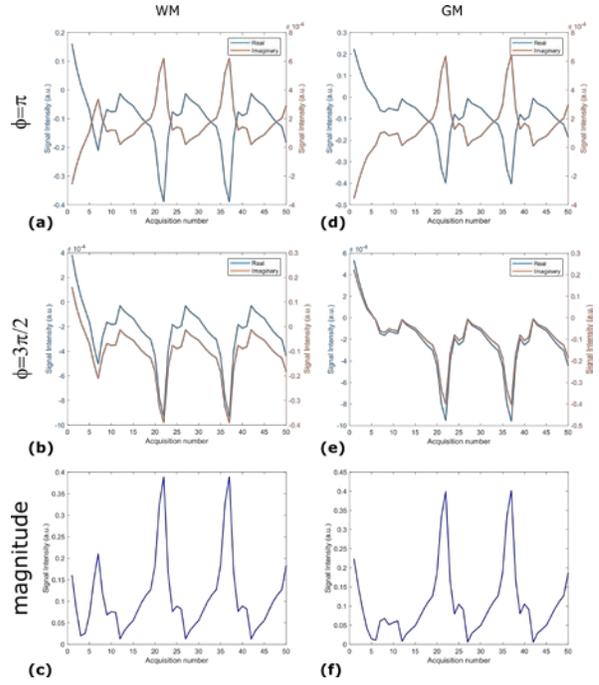

Fig. 4. Real and imaginary components of the simulated WM and GM signals for different phase values in comparison to the magnitude signal. (a) WM signal for Φ=π. (b) WM signal for Φ=3π/2. (c) WM magnitude signal. (d) GM signal for Φ=π. (e) GM signal for Φ=3π/2. (f) GM magnitude signal.

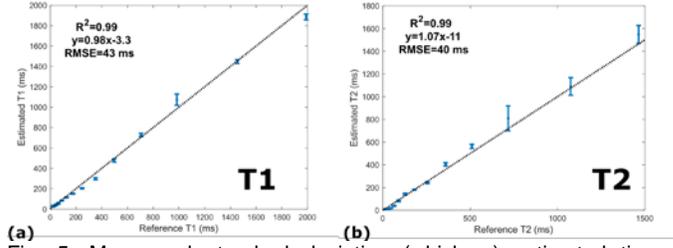

Fig. 5. Mean and standard deviation (whiskers) estimated tissue parameter values in each phantom compartment in comparison to the reference values. The correlation coefficient, RMSE values and equation for the best-fit line are shown inset. (a) MRF-EPI estimated T1 values in comparison to the reference values. (B) Estimated T2 values in comparison to reference values.

### C. In Vivo Healthy Subject

The reconstructed tissue parameter maps for the healthy subject are shown in Fig. 6. Reconstructions using the magnitude (Fig. 6a-d) and the complex data without phase estimation (Fig. 6e-h) resulted in significant artifacts or quantification errors, unlike the phase estimated complex reconstruction (Fig. 6i-m). The phase map (Fig. 6m) was shifted to the [-π,π] range by subtracting 2π from each voxel to facilitate visualization. The mean T1 and T2 values in WM and GM for each method are listed in Table I.

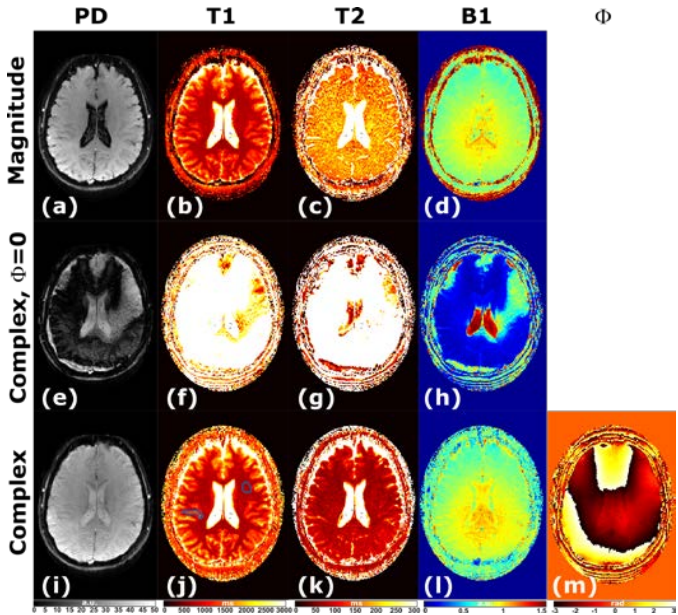

Fig. 6. Comparison of different DRONE quantifications. (a-d) Quantitative maps obtained using magnitude-only images resulted in erroneous T2 values. (e-h) Reconstruction using the complex data but without accounting for the variable phase resulted in severe errors and artifacts in all maps. (i-m) Reconstruction using the complex data and inclusion of the phase term Φ yielded anatomically correct maps and T1/T2 values like those obtained with other methods [31]. The areas outlined in blue in (j) indicate the WM and GM ROIs used to compare the values obtained with each reconstruction method.

The set of tissue parameter maps for a subset of 9 representative slices is shown in Fig. 7.

We manually segmented WM and deep GM brain structures and calculated the corresponding mean T1, T2 and $\chi$ values for each structure as shown in Table II.

### D. Patient with brain metastases

An example of tissue maps obtained in a 40-year-old male subject with metastatic melanoma is shown in Fig. 8. A subset of the lesions is marked with yellow arrows to illustrate the multifocal nature of brain metastases.

The distribution of the T1, T2 and $\chi$ values in each of the segmented tumor regions are shown in a box and whiskers plot with the median, first and third quartile ranges (Fig. 9). The differences in the tissue parameter values for each pair of segmented tumor regions was statistically significant ($p$=0.001) except as shown. Also shown are the tissue parameter values for an untreated tumor for comparison. Differences between the parameter values in the irradiated and untreated tumors were statistically significant as well.

## I. DISCUSSION

This work proposes to combine an optimized MRF-EPI acquisition, requiring less than 5 minutes, with a phase-sensitive deep learning quantification to obtain PD, T1, T2, B1 and $\chi$ maps. PS-DRONE provides accurate T1 and T2 maps (Fig. 5) and susceptibility maps that accurately reflect brain iron distribution (Fig. 7 and Table II).

There are important differences between the optimized MRF-EPI pulse sequence [11], [23] used here and others reported in the literature [8]–[10]. Other approaches utilized a variable TE in addition varying the flip angle and TR in each time-point. The resulting signal was thus sensitive to T2* but not T2. Here, the MRF-EPI pulse sequence used a constant TE, only varying the flip angles and the TR according to an optimized schedule. The signal variations for different tissues thus resulted from stimulated echoes refocused by subsequent RF pulses, effectively yielding a spin-echo-like sequence that was T2 sensitive. This is confirmed by the phantom results (Fig. 5), which showed good agreement between the reference values, obtained with spectroscopic spin-echo measurements, and the estimated values from the optimized MRF-EPI.

Recently, an MRF pulse sequence using quadratic-phase encoding was reported to yield T1, T2 and $\chi$ maps [33]. Because it was based on a spiral readout of k-space, scan time was over two times longer than the method proposed here. Additionally, the quantification required relatively large (3.2 GB) dictionaries and an additional refining step in the reconstruction. Finally, B1 variations were not accounted for.

TABLE I

COMPARISON OF WM AND GM RELAXATION VALUES FOR DIFFERENT DRONE RECONSTRUCTIONS

| | Reconstruction Type | | | | | | | |
|---|---|---|---|---|---|---|---|---|
| | Magnitude | | Complex with Φ=0 | | Complex | | Reference[a] | |
| Tissue | T1 (ms) | T2 (ms) | T1 (ms) | T2 (ms) | T1 (ms) | T2 (ms) | T1 (ms) | T2 (ms) |
| WM | 750.2±37.7 | 181.1±26.8 | 3994.3±396.7 | 3103.5±489.8 | 737.3±51.0 | 70.2±8.8 | 728±433 | 75±3 |
| GM | 1136.0±189.1 | 192.8±26.4 | 3927.2±328.6 | 1928.6±706.5 | 1265.5±225.3 | 75.5±11.4 | 1165±113 | 83±4 |

[a]Lu et al [31]

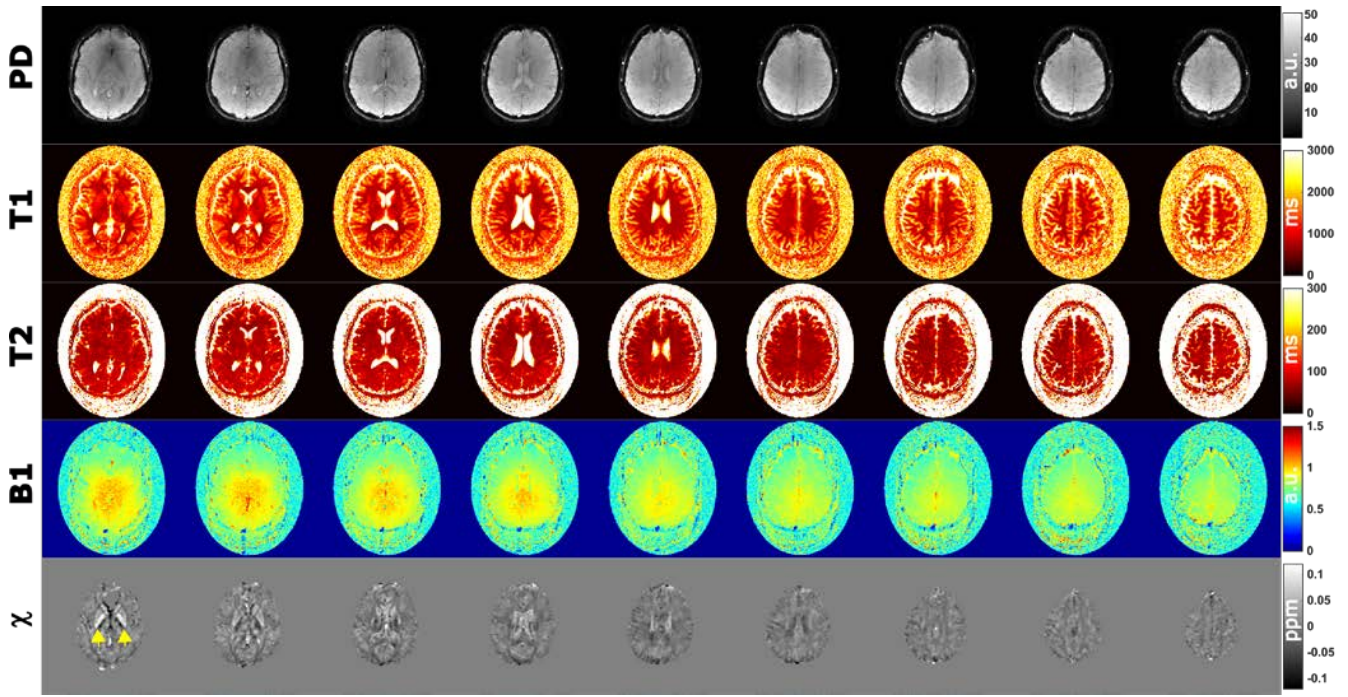

Fig. 7. Tissue maps for a subset of slices from the healthy subject acquisition. The PS-DRONE estimated phase was used to calculate the χ maps shown. Notable features include elevated susceptibility values in the globus pallidus (yellow arrows) resulting from increased iron deposition in that region.

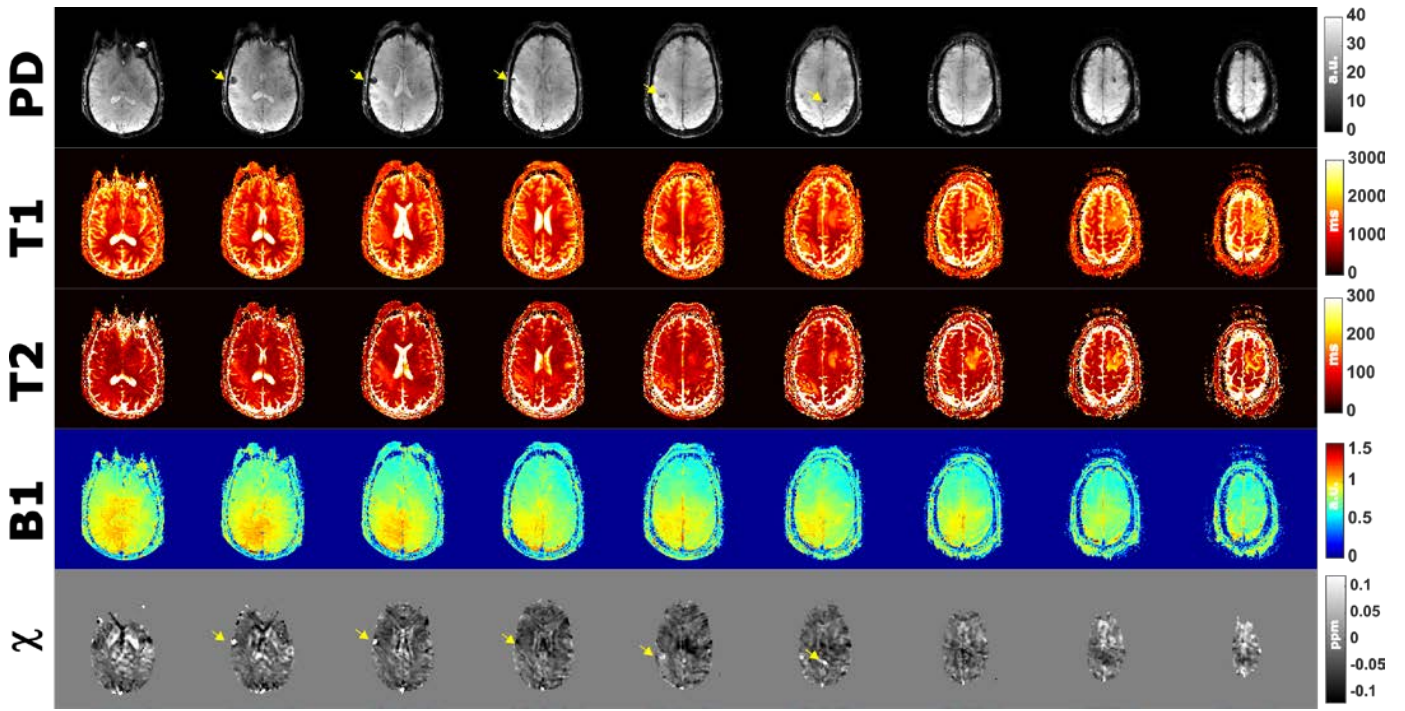

Fig. 8. Quantitative maps generated with PS-DRONE from a 40-year-old male subject with metastatic melanoma. Some of the multiple lesions in this subject are denoted with yellow arrows. Note the different manifestation of the tumor features in each map.

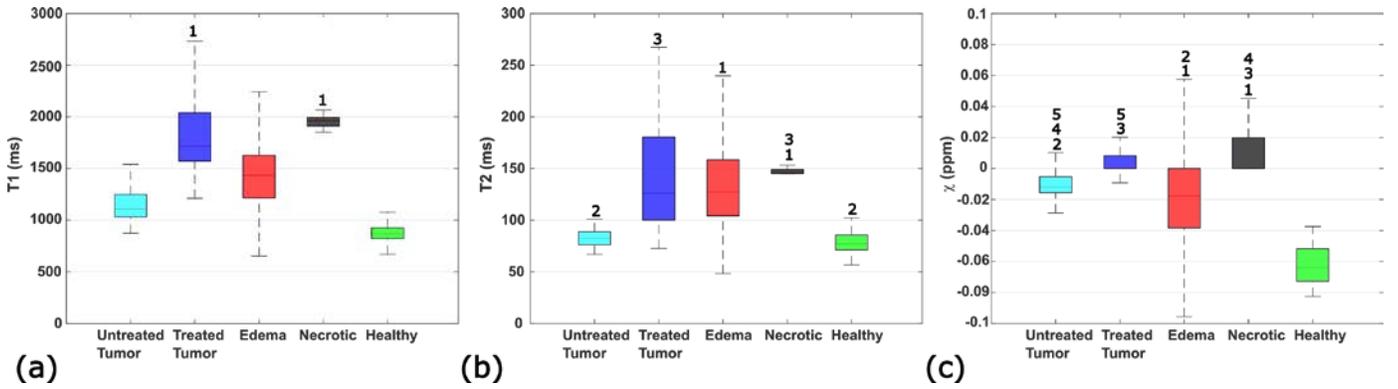

Fig. 9. Box and whiskers plot of the tissue parameter values in the tumor and contralateral ROIs. Shown are the distribution of the parameter values along with the median, first and third quartile ranges for (a) T1, (b) T2 and (c) χ maps. The significance of the differences between pairs of ROIs for each tissue parameter were calculated with a multi-comparison ANOVA test with Tukey HSD [32]. All pairs were statistically significant ($p$=0.001) except those denoted with corresponding digits above the plots. Note the significant differences between the irradiated and untreated tumors and between the tumor and healthy ROIs.

The use of PS-DRONE resolves many of the challenges inherent to dictionary matching. The network can be trained on a sparse dataset rather than requiring a dense sampling of the parameter space, as in dictionary matching, and the inclusion of additional parameters to the reconstruction is easily accomplished. Reconstruction time (approximately 1 second) is also significantly lower than dictionary matching.

The use of neural networks for multiparametric quantification is becoming increasingly common although most studies to date have used the magnitude signal alone [34]–[36]. At high fields, larger susceptibilities lead to increased phase dispersion and consequent errors in the reconstructed parameter maps if not properly accounted for. By jointly estimating the phase map along with the other parameters, PS-DRONE provides more accurate tissue maps (Fig. 6) while providing additional, clinically valuable maps of the susceptibility (Fig. 7-9).

Metastatic lesions are frequently multifocal [37] which necessitates whole brain imaging. Additionally, lesion presentation, particularly after radiation treatment, is often ambiguous hence requiring multiple and/or specialized scans to accurately characterize the lesion [38] and resulting in long (> 1 hour) scan times. The methods proposed here can alleviate this issue by providing rapid multiparametric maps that may offer improved diagnostic power given the statistically significant differences between healthy and diseased tissue, or between treated and untreated lesions as shown in Fig. 9. However, given the large biological variability in tumors, large-scale prospective studies will be required to verify this.

There are a several limitations to this work. Although the use of an EPI readout provides short acquisition times, EPI is also subject to well-known susceptibility artifacts that give rise to geometric distortions [39] which can affect the resulting tissue maps as well. This can be mitigated by switching to a multishot acquisition [40] albeit at the cost of increased scan time. Since scan time in our method is short (~3 minutes), and can be made even shorter with the inclusion of simultaneous multi-slice techniques [41], this further extension still has great promise in practical clinical implementation and will be explored in future work. The QSM pipeline used in this work is a for proof-of-concept and was not tailored or optimized for the data used. Indeed, the QSM tools and algorithms (e.g. MEDI) used were originally developed for multi-echo GRE acquisitions and not optimized for the MRF-EPI data acquired. Further optimization of the QSM pipeline will be explored in future studies as well.

TABLE II

TISSUE PARAMETER VALUES FOR VARIOUS BRAIN STRUCTURES OF A HEALTHY VOLUNTEER

| Brain Structure | Tissue Parameters | | |
| --- | --- | --- | --- |
| | T1 (ms) | T2 (ms) | χ (ppm) |
| Putamen | 1102.67±123.42 | 74.21±11.11 | 0.03±0.03 |
| Globus Pallidus | 901.26±111.91 | 71.51±12.60 | 0.09±0.02 |
| Head of caudate nucleus | 1225.76±118.29 | 81.07±10.28 | 0.05±0.01 |
| Genu of corpus callosum | 746.77±122.25 | 69.31±14.74 | -0.02±0.03 |

## II. CONCLUSION

A novel quantitative MRI framework was introduced to enable reconstruction of complex data using real-valued neural networks and improve the quantitative accuracy of the resulting parametric maps. In combination with an optimized MRF-EPI acquisition, this approach provides whole brain PD, T1, T2, B1 and χ maps in less than 5 minutes, which can facilitate clinical adoption. The maps obtained showed clear differences between healthy and diseased tissues and demonstrate the potential in assisting future clinical decision making.

## ACKNOWLEDGMENT

The authors are grateful to Dr. Youngwook Kee for his assistance with the QSM processing.

## CONFLICT OF INTEREST

OC holds patents for the DRONE and optimized MRF-EPI methods described in this work.

## REFERENCES

[1] V. Khoo, A. Padhani, S. Tanner, D. Finnigan, M. Leach, and D. Dearnaley, "Comparison of MRI with CT for the radiotherapy planning of prostate cancer: a feasibility study.," *Br. J. Radiol.*, vol. 72, no. 858, pp. 590–597, 1999.

[2] R. N. Low, R. M. Barone, and J. Lucero, "Comparison of MRI and CT for predicting the Peritoneal Cancer Index (PCI) preoperatively in patients being considered for cytoreductive surgical procedures," *Ann. Surg. Oncol.*, vol. 22, no. 5, pp. 1708–1715, 2015.

[3] R. Ordidge, P. Gibbs, B. Chapman, M. Stehling, and P. Mansfield, "High-speed multislice T1 mapping using inversion-recovery echo-planar imaging," *Magn. Reson. Med.*, vol. 16, no. 2, pp. 238–245, 1990.

[4] R. Darwin, B. Drayer, S. Riederer, H. Wang, and J. MacFall, "T2 estimates in healthy and diseased brain tissue: a comparison using various MR pulse sequences.," *Radiology*, vol. 160, no. 2, pp. 375–381, 1986.

[5] D. Ma et al., "Magnetic resonance fingerprinting," *Nature*, vol. 495, no. 7440, pp. 187–192, 2013.

[6] Y. Jiang, D. Ma, N. Seiberlich, V. Gulani, and M. A. Griswold, "MR fingerprinting using fast imaging with steady state precession (FISP) with spiral readout," *Magn. Reson. Med.*, vol. 74, no. 6, pp. 1621–1631, 2015.

[7] M. A. Cloos et al., "Rapid radial T1 and T2 mapping of the hip articular cartilage with magnetic resonance fingerprinting," *J. Magn. Reson. Imaging*, vol. 50, no. 3, pp. 810–815, 2019.

[8] I. Hermann et al., "Accelerated white matter lesion analysis based on simultaneous T1 and T2∗ quantification using magnetic resonance fingerprinting and deep learning," *Magn. Reson. Med.*, vol. 86, no. 1, pp. 471–486, 2021.

[9] B. Rieger et al., "Time efficient whole-brain coverage with MR Fingerprinting using slice-interleaved echo-planar-imaging," *Sci. Rep.*, vol. 8, no. 1, pp. 1–12, 2018.

[10] B. Rieger, F. Zimmer, J. Zapp, S. Weingärtner, and L. R. Schad, "Magnetic resonance fingerprinting using echo-planar imaging: Joint quantification of T1 and relaxation times," *Magn. Reson. Med.*, vol. 78, no. 5, pp. 1724–1733, 2017.

[11] O. Cohen, M. Sarracanie, M. S. Rosen, and J. L. Ackerman, "In vivo optimized MR fingerprinting in the human brain," in *Proceedings of the International Society of Magnetic Resonance in Medicine*, 2016, p. 0430.

[12] O. Cohen, B. Zhu, and M. S. Rosen, "MR fingerprinting deep reconstruction network (DRONE)," *Magn. Reson. Med.*, vol. 80, no. 3, pp. 885–894, 2018.

[13] N. Stikov, M. Boudreau, I. R. Levesque, C. L. Tardif, J. K. Barral, and G. B. Pike, "On the accuracy of T1 mapping: searching for common ground," *Magn. Reson. Med.*, vol. 73, no. 2, pp. 514–522, 2015.

[14] P. Virtue, X. Y. Stella, and M. Lustig, "Better than real: Complex-valued neural nets for MRI fingerprinting," in *2017 IEEE international conference on image processing (ICIP)*, 2017, pp. 3953–3957.

[15] A. Deistung, F. Schweser, and J. R. Reichenbach, "Overview of quantitative susceptibility mapping," *NMR Biomed.*, vol. 30, no. 4, p. e3569, 2017.

[16] Y. Wang and T. Liu, "Quantitative susceptibility mapping (QSM): decoding MRI data for a tissue magnetic biomarker," *Magn. Reson. Med.*, vol. 73, no. 1, pp. 82–101, 2015.

[17] C. Bakker, C. De Graaf, and P. Van Dijk, "Restoration of signal polarity in a set of inversion recovery NMR images," *IEEE Trans. Med. Imaging*, vol. 3, no. 4, pp. 197–202, 1984.

[18] P. Hou, K. M. Hasan, C. W. Sitton, J. S. Wolinsky, and P. A. Narayana, "Phase-sensitive T1 inversion recovery imaging: a time-efficient interleaved technique for improved tissue contrast in neuroimaging," *Am. J. Neuroradiol.*, vol. 26, no. 6, pp. 1432–1438, 2005.

[19] J. A. Borrello, T. L. Chenevert, and A. M. Aisen, "Regional phase correction of inversion-recovery MR images," *Magn. Reson. Med.*, vol. 14, no. 1, pp. 56–67, 1990.

[20] K. F. Stupic et al., "A standard system phantom for magnetic resonance imaging," *Magn. Reson. Med.*, 2021.

[21] F. Bloch, "Nuclear induction," *Phys. Rev.*, vol. 70, no. 7–8, p. 460, 1946.

[22] M. Weigel, "Extended phase graphs: dephasing, RF pulses, and echoes-pure and simple," *J. Magn. Reson. Imaging*, vol. 41, no. 2, pp. 266–295, 2015.

[23] O. Cohen and M. S. Rosen, "Algorithm comparison for schedule optimization in MR fingerprinting," *Magn. Reson. Imaging*, vol. 41, pp. 15–21, 2017.

[24] A. Paszke et al., "Pytorch: An imperative style, high-performance deep learning library," *Adv. Neural Inf. Process. Syst.*, vol. 32, pp. 8026–8037, 2019.

[25] J. Z. Bojorquez, S. Bricq, C. Acquitter, F. Brunotte, P. M. Walker, and A. Lalande, "What are normal relaxation times of tissues at 3 T?," *Magn. Reson. Imaging*, vol. 35, pp. 69–80, 2017.

[26] M. A. Herráez, D. R. Burton, M. J. Lalor, and M. A. Gdeisat, "Fast two-dimensional phase-unwrapping algorithm based on sorting by reliability following a noncontinuous path," *Appl. Opt.*, vol. 41, no. 35, pp. 7437–7444, 2002.

[27] D. Zhou, T. Liu, P. Spincemaille, and Y. Wang, "Background field removal by solving the Laplacian boundary value problem," *NMR Biomed.*, vol. 27, no. 3, pp. 312–319, 2014.

[28] T. Liu, C. Wisnieff, M. Lou, W. Chen, P. Spincemaille, and Y. Wang, "Nonlinear formulation of the magnetic field to source relationship for robust quantitative susceptibility mapping," *Magn. Reson. Med.*, vol. 69, no. 2, pp. 467–476, 2013.

[29] T. Liu et al., "Morphology enabled dipole inversion (MEDI) from a single-angle acquisition: comparison with COSMOS in human brain imaging," *Magn. Reson. Med.*, vol. 66, no. 3, pp. 777–783, 2011.

[30] L. De Rochefort et al., "Quantitative susceptibility map reconstruction from MR phase data using bayesian regularization: validation and application to brain imaging," *Magn. Reson. Med. Off. J. Int. Soc. Magn. Reson. Med.*, vol. 63, no. 1, pp. 194–206, 2010.

[31] H. Lu, L. M. Nagae-Poetscher, X. Golay, D. Lin, M. Pomper, and P. C. Van Zijl, "Routine clinical brain MRI sequences for use at 3.0 Tesla," *J. Magn. Reson. Imaging Off. J. Int. Soc. Magn. Reson. Med.*, vol. 22, no. 1, pp. 13–22, 2005.

[32] J. W. Tukey, "Comparing individual means in the analysis of variance," *Biometrics*, pp. 99–114, 1949.

[33] R. Boyacioglu, C. Wang, D. Ma, D. F. McGivney, X. Yu, and M. A. Griswold, "3D magnetic resonance fingerprinting with quadratic RF phase," *Magn. Reson. Med.*, vol. 85, no. 4, pp. 2084–2094, 2021.

[34] O. Perlman et al., "Quantitative imaging of apoptosis following oncolytic virotherapy by magnetic resonance fingerprinting aided by deep learning," *Nat. Biomed. Eng.*, 2021, doi: 10.1038/s41551-021-00809-7.

[35] F. Glang et al., "DeepCEST 3T: Robust MRI parameter determination and uncertainty quantification with neural networks—application to CEST imaging of the human brain at 3T," *Magn. Reson. Med.*, vol. 84, no. 1, pp. 450–466, 2020.

[36] Q. Zhang et al., "Deep learning–based MR fingerprinting ASL ReconStruction (DeepMARS)," *Magn. Reson. Med.*, vol. 84, no. 2, pp. 1024–1034, 2020.

[37] E. S. Nussbaum, H. R. Djalilian, K. H. Cho, and W. A. Hall, "Brain metastases: histology, multiplicity, surgery, and survival," *Cancer Interdiscip. Int. J. Am. Cancer Soc.*, vol. 78, no. 8, pp. 1781–1788, 1996.

[38] K. Parvez, A. Parvez, and G. Zadeh, "The diagnosis and treatment of pseudoprogression, radiation necrosis and brain tumor recurrence," *Int. J. Mol. Sci.*, vol. 15, no. 7, pp. 11832–11846, 2014.

[39] P. Jezzard and R. S. Balaban, "Correction for geometric distortion in echo planar images from B0 field variations," *Magn. Reson. Med.*, vol. 34, no. 1, pp. 65–73, 1995.

[40] N. Chen, A. Guidon, H.-C. Chang, and A. W. Song, "A robust multi-shot scan strategy for high-resolution diffusion weighted MRI enabled by multiplexed sensitivity-encoding (MUSE)," *Neuroimage*, vol. 72, pp. 41–47, 2013.

[41] D. A. Feinberg and K. Setsompop, "Ultra-fast MRI of the human brain with simultaneous multi-slice imaging," *J. Magn. Reson.*, vol. 229, pp. 90–100, 2013.